\documentclass{article}
\pdfoutput=1
\usepackage{amsmath,amssymb,epsfig,color}
\numberwithin{equation}{section}
\usepackage{hyperref, cite}

\title{Von Neumann entropy and Lindblad decoherence\\ in the high energy limit of strong interactions} 
\author{G. Chachamis$^{1,2}$, M. Hentschinski$^3$, A. Sabio Vera$^{4,5}$\\ \\
\small $^1$ Laborat{\' o}rio de Instrumenta\c{c}{\~ a}o e F{\' \i}sica Experimental de Part{\' \i}culas (LIP),\\
\small Av. Prof. Gama Pinto, 2, P-1649-003 Lisboa, Portugal.\\
\small $^2$Departamento de Estad{\'\i}stica, Inform{\'a}tica y Matem{\'a}ticas, \\
\small Universidad P{\'u}blica de Navarra--UPNA, 31006 Pamplona, Spain.\\
\small $^3$ Departamento de Actuaria, F{\'\i}sica y Matem{\'a}ticas, Universidad de las Americas Puebla, \\
\small Santa Catarina Martir, San Andr{\'e}s Cholula, 72820 Puebla, Mexico.\\
\small $^4$ Theoretical Physics Department, Universidad Aut{\' o}noma de Madrid, Madrid 28049, Spain.\\
\small $^5$ Instituto de F{\'\i}sica Te{\' o}rica UAM/CSIC, c/ Nicol{\' a}s Cabrera 13-15, \\
\small Universidad Aut{\' o}noma de Madrid Cantoblanco, Madrid 28049, Spain.}

\begin{document} 

\maketitle 

\begin{abstract}
Quantum properties of the state associated to the gluon Green's function in the BFKL approach are studied using a  discretization in  virtuality space. Considering the coupling constant as imaginary, its density matrix corresponds to a pure state for any energy. Non--linear corrections due to high gluon densities are modelled through a suppression of infrared modes in the Hamiltonian making it no longer hermitian.  This introduces quantum decoherence into the evolution equation. When the coupling is real this leads to unbounded normalization of states which becomes  bounded for sufficient saturation of infrared modes. Physical  quantum properties,  such as a purity smaller than one or a positive von Neumann entropy, hence are recovered when the infrared/ultraviolet original symmetry of the formalism is broken. Similarly to the work of Armesto, Dom{\'\i}nguez, Kovner, Lublinsky and Skokov in~\cite{Armesto:2019mna}, an evolution equation of Lindblad type for the normalized density matrix describing the open system is obtained. 
\end{abstract}

\section{Introduction}
\label{sec:intro}

Application of concepts of quantum information in the context of hadronic reactions and the microscopic theory of strong interactions,  Quantum Chromodynamics (QCD), has been a field of intense research during recent years. At the heart of this interest lies the possible relation of the confinement problem of strong interactions with  entanglement: Confinement of colored charges into colorless hadrons is interpreted as a particularly strong form of entanglement of microscopic degrees of freedom~\cite{Klebanov:2007ws,Kharzeev:2017qzs,Beane:2018oxh,Mueller:2019qqj,Lamm:2019uyc,Chakraborty:2020uhf,Liu:2020eoa,Briceno:2020rar,Davoudi:2020yln,deJong:2021wsd,Barata:2021yri,Li:2021kcs,Gong:2021bcp}, see Ref.~\cite{Beck:2023xhh} for a recent review. Several proposals \cite{Dumitru:2023qee,Duan:2021clk,Dumitru:2022tud,Ramos:2022gia,Kou:2022dkw,Ehlers:2022oal,Asadi:2023bat,Kou:2023azd,Kutak:2023cwg,Gursoy:2023hge,Barata:2023jgd,Hentschinski:2023izh}  have emerged in the literature that address different ways to study entanglement in the context of hadronic reactions. 

An interesting subset of such studies refers to the creation of entanglement entropy in the so--called low $x$ limit of QCD. With $Q^2$ being the resolution scale in a Deep Inelastic Scattering (DIS) event,  Bjorken $x$ is generically defined as the ratio of this hard scale and the squared center--of--mass energy $s$. The low $x$ limit therefore refers to the perturbative high energy limit of strong interactions. Recent studies \cite{Hentschinski:2023izh,Hentschinski:2022rsa,Hentschinski:2021aux, Liu:2023eve,Liu:2023zno,Liu:2022bru,Liu:2022qqf,Liu:2022hto,Liu:2022ohy,Liu:2022urb,Liu:2019yye} explore entanglement and its imprints in multiplicity distributions in the low $x$ limit using the color dipole model, where the evolution towards large  $Y=\ln(1/x)$ is understood as the subsequent branching of color dipoles (see also \cite{Armesto:2019mna, Duan:2020jkz} for studies within the related Color Glass Condensate framework). 

In the present work we take a slighly different perspective. Instead of
making use of the color dipole picture, we study the QCD density
matrix in the low $x$ limit within the Balitsky--Fadin--Kuraev--Lipatov (BFKL) formalism\cite{Lipatov:1985uk,Lipatov:1976zz,Fadin:1975cb,Kuraev:1976ge,Kuraev:1977fs,Balitsky:1978ic}. This
framework identifies reggeized gluons as the relevant degrees of
freedom in the $t$-channel of high energy factorized scattering
amplitudes, which form the starting point for a resummation of high
energy logarithms (see~\cite{Lipatov:1995pn, Lipatov:1996ts,
  Hentschinski:2018rrf,Hentschinski:2011tz,GomezBock:2020zxp,Hentschinski:2020rfx,Hentschinski:2020tbi,Hentschinski:2021lsh}
for a discussion and derivation of the BFKL evolution in the context of an
effective action framework, based on reggeized gluon fields).  For
perturbative scattering amplitudes, high energy factorization is
achieved via the exchange of a single reggeized gluon which
results at cross-section level into a two reggeized gluon state in the
overall color singlet state. This constitutes the starting point for the resummation of
perturbative terms enhanced by powers of $Y$ to all orders in the
strong coupling constant, generating a bound state of  two
reggeized gluons called the hard or BFKL Pomeron (see~\cite{Chachamis:2022jis} for a recent work on the conformal properties of this bound state).

 This procedure generates a powerlike rise of cross-sections with $s$. While such a rise has been observed in experimental data (see, {\it e.g.}~\cite{Hentschinski:2012kr}),  it eventually leads -- if continued to
arbitarily high center--of--mass energies -- to a violation of unitary
bounds. To tame this growth, it is hence needed to extend the resummation to
the exchange of multiple reggeized gluons allowing for vertices changing the number of
exchanged reggeized gluons in the $t$-channel. For a sufficiently inclusive cross-section, the simplest
one is the 2 to 4 reggeized gluon transition vertex which in
the multicolor limit turns into a triple Pomeron vertex. Apart from
slowing down the growth with energy, inclusion of such number--changing
elements  also has an important consequence in the dynamics of the 
transverse momentum space.  While the BFKL kernel is symmetric with respect
to incoming and outgoing $t$--channel momenta, this symmetry is broken
once number--changing elements, such as the triple Pomeron vertex are
included\cite{Mueller:2002zm, Bartels:2007dm}. The combination of  linear BFKL 
evolution with these new vertices results in a 
cancellation of infrared modes and the evolution acquires an effective scale,
known as saturation scale~\cite{Gribov:1983ivg,McLerran:1993ni}, which increases with $Y$.

In the following we explore the consequences of such dynamics using an explicit construction of 
a density matrix for the two reggeized
gluon state, employing a matrix representation of the leading order
BFKL evolution equation first proposed in~\cite{BethencourtdeLeon:2011xks}, which stems from a 
discretization of the dynamics in transverse momentum space. This approach is useful because 
it allows a transparent understanding of quantum mechanical properties of the scattering process.

The outline of our work is as follows: In Sec.~\ref{sec:hamilt-matr-repr} we provide an introduction to the matrix representation of BFKL evolution, while in Sec.~\ref{sec:dens-matr-virt} the corresponding density matrix is investigated.  Sec.~\ref{sec:scre-infr-diff} introduces a modification of this framework due to infrared screening, while in Sec. \ref{sec:non-herm-hamilt} we explore the consequences of the resulting non--hermitian Hamiltonian and derive an evolution equation of the Lindblad type. In Sec.~\ref{sec:conclusions} we finally present the  conclusions and outlook for future work.  

\section{Hamiltonian in matrix representation}
\label{sec:hamilt-matr-repr}

High energy scattering in QCD and supersymmetric theories can be described by the BFKL approach when the leading logarithms of the center--of--mass energy are resummed~\cite{Lipatov:1985uk,Lipatov:1976zz,Fadin:1975cb,Kuraev:1976ge,Kuraev:1977fs,Balitsky:1978ic}. Following the work in Ref.~\cite{BethencourtdeLeon:2011xks} we consider the azimuthal--angle averaged forward BFKL equation and discretize it in the virtuality space 
of $t$--channel Reggeized gluons. After regularizing it in a finite--length box we obtain a square--matrix representation of the Hamiltonian. Its spectrum contains positive and negative eigenvalues~\cite{BethencourtdeLeon:2011xks}. If the virtuality space for the propagators of Reggeized gluons   in the forward BFKL equation is discretized, after azimuthal angle averaging, the following matrix representation is obtained
\begin{eqnarray}
\frac{\partial }{\alpha \partial Y} \left|{\phi}^{(N)}\right>
&=& \hat{{\cal H}}^{{\Box }}_N  \, \left|{\phi}^{(N)} \right> \, ,
\label{BFKLeqn1}
\end{eqnarray}
where $\alpha = \alpha_s N_c/\pi$, see ~\cite{BethencourtdeLeon:2011xks} for the details of this result. Note that each iteration of this equation corresponds to a contribution to the total cross section, structure functions in DIS. The Hamiltonian matrix elements take the form
\begin{eqnarray}
(\hat{\mathcal{H}}^{{\Box}}_N)_{i, j} &=& \sum_{n=1}^{N-1} \frac{\delta_i^{j+n}}{n}+\sum_{n=1}^{N-1} \frac{\delta_{i+n}^j}{n}-2 h(i-1) \delta_i^j \nonumber\\
&=& \frac{\theta(i-j)}{i -j} + \frac{\theta(j-i)}{j -i} - 2 h(i-1) \delta_i^j\, .
\label{BFKLHamilt1}
\end{eqnarray}
$h(i)$ is the harmonic number. To obtain this square $N \times N$ matrix representation  an upper cut--off in the virtuality integrations is introduced. This corresponds to a one--dimensional box. The limit to recover the original equation corresponds to  $N \to \infty$. Eq.~\ref{BFKLHamilt1} is built from shift matrices, 
\begin{eqnarray}
{\cal H}^{\Box }_N &=& \sum_{n=1}^{N-1} \frac{\left(\hat{\cal S}_{\rm IR}\right)^n}{n}
+\sum_{n=1}^{N-1} \frac{\left(\hat{\cal S}_{\rm UV}\right)^n}{n} + \hat{\cal G},
\end{eqnarray}
where $(\hat{\cal G})_{i,j}= -2 h(i-1) \delta_i^j$, $(\hat{\cal S}_{\rm IR})_{i,j} = \delta_{i}^{j+1}$ and $(\hat{\cal S}_{\rm UV})_{i,j} = \delta_{i+1}^{j}$. This is natural since the Hamiltonian generates the symmetric diffusion towards infrared and ultraviolet values of the virtuality starting at the initial condition, towards the quantum state at a generic $Y$. These are driven, respectively, by $\hat{\cal S}_{\rm IR}$ and 
$\hat{\cal S}_{\rm UV}$. $\hat{\cal G}$ accounts for Reggeized $t$--channel propagators, which correspond to the generation of multiple rapidity gaps in the final state.

\begin{figure}
\center
  \includegraphics[width=10.cm]{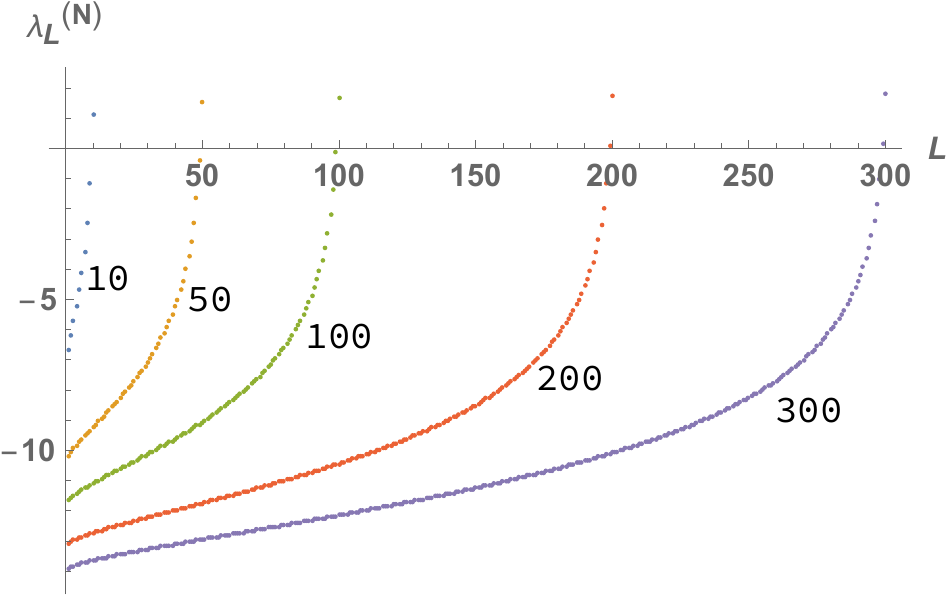}
  \caption{Set of eigenvalues for the $N \times N$ Hamiltonian with $N=10,50,100,200,300$. }
  \label{fig:EigenvaluesUpto300}
\end{figure}

In order to operate in a Hilbert space of normalized quantum states at any value of the energy, an analytic continuation in the coupling constant to the line (with real $\lambda$) $\alpha = i \lambda$ 
is performed. A standard Schr{\"o}dinger equation then drives the dynamics:
\begin{eqnarray}
i\frac{\partial }{ \partial Y} \left|{\phi}^{(N)}\right>
&=& -\lambda \hat{{\cal H}}^{{\Box }}_N  \, \left|{\phi}^{(N)} \right> \, .
\label{BFKLeqn2}
\end{eqnarray}
The formal solution can be written in iterative form, 
\begin{eqnarray}
\left|{\phi}^{(N)} \right>&=& e^{ i \lambda Y \hat{\mathcal{H}}^{{\Box }}_N} \,  \left|{\varphi}_0^{(N_0)} \right>\nonumber\\
&&\hspace{-1cm}=~ \left\{1+\int_0^Y d y_1\left( i \lambda \hat{\mathcal{H}}^{{\Box }}_N\right) + \int_0^Y d y_1\left( i \lambda \hat{\mathcal{H}}^{{\Box }}_N\right) \int_0^{y_1} d y_2\left( i \lambda \hat{\mathcal{H}}^{ {\Box }}_N\right)  \right.\nonumber\\
&&\hspace{-1cm}+~ \left. \int_0^Y d y_1\left( i \lambda \hat{\mathcal{H}}^{{\Box }}_N\right) \int_0^{y_1} d y_2\left( i \lambda \hat{\mathcal{H}}^{{\Box }}_N\right)  \int_0^{y_2} d y_3\left( i \lambda \hat{\mathcal{H}}^{{\Box }}_N\right)+~ \cdots \right\} \,  \left|{\varphi}_0^{(N_0)}   \right>
\, ,
\end{eqnarray}
where the initial condition $\left|{\varphi}_0^{(N_0)}  \right> \equiv\left(\varphi_1^0, \varphi_2^0, \ldots, \varphi_N^0\right)^T$ excites a single virtuality component:  $\varphi_i^0=\delta_i^{N_0}$ and corresponds to the square of a single gluon propagator in the $t$--channel, the tree level contribution to the scattering amplitude in the forward limit.  
If $\left|{\phi}^{(N)} \right>= \left(\phi_1, \phi_2, \ldots, \phi_N\right)^T$  then 
\begin{eqnarray}
i \frac{\partial \phi_j}{ \partial Y} &=& - \lambda \sum_{l=1}^N\left(\frac{\left(1-\delta_l^j\right)}{|l-j|}-2 h(j-1) \delta_l^j\right) \phi_l \, .
\end{eqnarray}
The growth with energy in this square truncation depends on the matrix size. The $N \times N$ matrix $\hat{{\cal H}}^{{\Box }}_N$ is symmetric and real. It can be diagonalized, it has normalized real eigenvectors $\left|{\psi}_L^{(N)}\right>$  with real eigenvalues  $\lambda_L^{(N)}$ ($L=1, \dots, N$), and spectral decomposition
\begin{eqnarray}
\hat{{\cal H}}^{{\Box }}_N &=& \sum_{L=1}^N \lambda_L^{(N)} \left|{\psi}_L^{(N)} \right> \left<{\psi}_L^{(N)} \right| 
\, .
\end{eqnarray}
 As it is shown in Fig.~\ref{fig:EigenvaluesUpto300}, the spectrum of $\hat{{\cal H}}^{{\Box }}_N$ has a largest positive eigenvalue  for any $N$, which tends to $4 \ln{2}$ when $N \to \infty$, with a gap with respect to the lower ones which are mostly negative~\cite{BethencourtdeLeon:2011xks}.  
The number of positive eigenvalues slowly grows with $N$ ({\it e.g.} the second positive eigenvalue appears when $N=165$). 

Since any initial condition vector may be expanded in the complete basis of eigenvectors, 
\begin{eqnarray}
\left|{\varphi}_0^{(N_0)} \right>&=& \sum_{L=1}^N c_L^{(N_0)} 
\left|{\psi}_L^{(N)} \right>\, ,
\label{InitialCondPureState}
\end{eqnarray}
 it is then possible to express the gluon Green's function state as
\begin{eqnarray}
\left|{\phi}^{(N)} \right>&=&e^{ i \lambda Y \hat{\mathcal{H}}^{{\Box }}_N}  \left|{\varphi}_0^{(N_0)} \right>~=~ \sum_{L=1}^N c_L^{(N_0)} e^{ i \lambda Y \lambda_L^{(N)}} \left|{\psi}_L^{~(N)} \right>\, ,
\label{DisGGF}
\end{eqnarray}
where, for any $Y$, $\left<{\phi}^{(N)} \right.\left|{\phi}^{(N)} \right> =1$  and  $\left< {\varphi}_0^{(N_0)} \right. \left|{\varphi}_0^{(N_0)} \right> = \sum_{L=1}^N |c_L^{(N_0)} |^2 ~=~ 1$.

For the sake of clarity, let us focus on the $N=5$ case with Hamiltonian
\begin{eqnarray}
\hat{\mathcal{H}}^{{\Box }}_{5} = \left(
\begin{array}{ccccc}
 0 & 1 & \frac{1}{2} & \frac{1}{3} & \frac{1}{4} \\
 1 & -2 & 1 & \frac{1}{2} & \frac{1}{3} \\
 \frac{1}{2} & 1 & -3 & 1 & \frac{1}{2} \\
 \frac{1}{3} & \frac{1}{2} & 1 & -\frac{11}{3} & 1 \\
 \frac{1}{4} & \frac{1}{3} & \frac{1}{2} & 1 & -\frac{25}{6} \\
\end{array}
\right) \, .
\end{eqnarray}
It has the eigenvalues $\left(\lambda_1^{(5)}, \dots , \lambda_{5}^{(5)} \right) = \left(-4.9838,-4.07483,-3.03006,-1.59174,0.847101\right)$. 
Since $ \left<{\psi}_L^{(5)} \right|\left.{\psi}_M^{(5)} \right>   = \delta_{L,M}$ then, for a particular initial condition, {\it e.g.} 
\begin{eqnarray}
\left|{\varphi}_0^{(3)} \right>&=& \left(
\begin{array}{c}
 0  \\
 0  \\
 1 \\
 0  \\
 0  \\
\end{array}
\right) ~=~ \sum_{L=1}^{5} c_L^{(3)}
\left| {\psi}_L^{(5)} \right>\, ,
 \label{InitiCond3}
 \end{eqnarray} 
 we have $\left(c_1^{(3)}, \dots , c_{5}^{(3)} \right) = 
 \left(0.166387, 0.753377, 0.278794, 0.492161, -0.291187\right)$.
The discretized version of the Green's function reads as in 
Eq.~\ref{DisGGF} (with $\eta=i \lambda Y$):
 \begin{eqnarray}
\left|{\phi}^{(5)} \right> \simeq 
\left(
\begin{array}{ccccc}
0.24 &  -0.28 &0.031 &0.014 &-0.0006 \\
0.13 &0.23 &-0.20 &-0.15 & -0.0040 \\
 0.08
   &0.24 &0.08 &0.6 &0.028\\
 0.06 &0.19 &0.14 &-0.27 &-0.11 \\
0.041 &0.12 &0.10 &-0.38 & 0.12 \\
\end{array}
\right)
\left(\begin{array}{c}
e^{0.85 \eta } \\
e^{-1.6 \eta}\\
e^{-3.0 \eta } \\
 e^{-4.1 \eta } \\
  e^{-5.0 \eta } \\
\end{array}\right).
 \end{eqnarray}

In the following section we investigate different aspects of this quantum system encoded in its density matrix which is suited for the description not only of  pure but mainly of mixed states.  

\section{Density matrix in virtuality space}
\label{sec:dens-matr-virt}

The vector state $\left|\phi^{(N)}\right>$ represents a system isolated from any external information which evolves with energy following~\ref{BFKLeqn2}. Unitary evolution  is driven by the  operator ${\hat U} (Y) = e^{i \lambda Y \hat{\cal H}_N^\square}$.
In the $N$--dimensional Hilbert space of discretized virtualities there exists a pure--state operator describing the initial condition~\ref{InitialCondPureState} at $Y=0$; $\hat{\rho}_{\rm pure}^{(N,N_0)} (0) =  \left|{\varphi}_0^{(N_0)}  \right>   \left<{\varphi}_0^{(N_0)}  \right| $. 

Its evolution with $Y$ is
\begin{eqnarray}
\hat{\rho}_{\rm pure}^{(N,N_0)} (Y) =
\left|{\phi}^{(N)} (Y)\right> \left<{\phi}^{(N)} (Y)\right| = \sum_{L,M=1}^N  
(\hat{\rho}_{\rm pure}^{(N,N_0)}(Y))_{L,M}   \left|{\psi}_L^{(N)} \right> \left<{\psi}_M^{(N)} \right| \, .
\label{Purestateoperator}
\end{eqnarray}
This is a (real) hermitian matrix with elements  $(\hat{\rho}_{\rm pure}^{(N,N_0)}(Y))_{L,M}  = 
c_L^{(N_0)} (c_M^{(N_0)})^*  e^{i \lambda Y (\lambda_L^{(N)} - \lambda_M^{(N)})}$.  For an infinitesimal energy interval $d Y$, 
\begin{eqnarray}
\hat{\rho}_{\rm pure}^{(N,N_0)} (Y+d Y) 
&\simeq&  
 \hat{\rho}_{\rm pure}^{(N,N_0)}(Y)+ i \lambda d Y \, \left[\hat{\cal H}_N^\square, \hat{\rho}_{\rm pure}^{(N,N_0)} (Y) \right] \, .
\end{eqnarray}
Therefore
\begin{eqnarray}
\frac{d \hat{\rho}_{\rm pure}^{(N,N_0)} }{i \lambda d Y} &=&\left[\hat{\cal H}_N^\square, \hat{\rho}_{\rm pure}^{(N,N_0)} \right] \nonumber\\
&=&  \sum_{L,M=1}^N  
c_L^{(N_0)} (c_M^{(N_0)})^*  (\lambda_L^{(N)} - \lambda_M^{(N)}) 
e^{i \lambda Y (\lambda_L^{(N)} - \lambda_M^{(N)})}    \left|{\psi}_L^{(N)} \right> \left<{\psi}_M^{(N)} \right|\, ,
\end{eqnarray}
a Liouville--von Neumann equation for the pure--density state operator~\ref{Purestateoperator}. Its trace is unity and time independent,
\begin{eqnarray}
{\rm Tr} \, (\hat{\rho}_{\rm pure}^{(N,N_0)} (Y)) =  
  \sum_{L,M=1}^N  
(\hat{\rho}_{\rm pure}^{(N,N_0)}(Y))_{L,M}    \left<{\psi}_M^{(N)} \right. \left|{\psi}_L^{(N)} \right> =   \sum_{L=1}^N  \left|c_L^{(N_0)}\right|^2  
= 1 \, .
\label{TracePurestateoperator}
\end{eqnarray}
This implies that $\hat{\rho}_{\rm pure}^{(N,N_0)}$ allows for the proper evaluation of expectation values of operators, 
$\small<\hspace{-.1cm}\hat{A}\hspace{-.1cm}\small>_Y = {\rm Tr} (\hat{A}\hat{\rho}_{\rm pure}^{(N,N_0)})$.  Since it is an idempotent matrix with trace one, it has a single non--zero eigenvalue $\lambda_{\hat{\rho}_{\rm pure}^{(N,N_0)}} = 1$. It is a projector onto a one-dimensional subspace within the Hilbert space of possible quantum states. There exists a complete knowledge of the state of the system at any $Y$. This density matrix is called a {\sl pure state}; its purity is one: ${\rm Tr} (\hat{\rho}_{\rm pure}^{(N,N_0)})^2 =1$. 

The fact that the rank of the density matrix associated to the Green's function state is one for any value of $Y$ implies that its {\sl von Neumann entropy},
\begin{eqnarray}
{\cal S}_{\rm vN}^{(N,N_0)} (Y) = - {\rm Tr} \left(\hat{\rho}_{\rm pure}^{(N,N_0)} (Y) \log_2{\hat{\rho}_{\rm pure}^{(N,N_0)} (Y)}\right) = - \lambda_{\hat{\rho}_{\rm pure}^{(N,N_0)}} (Y)   \log_2{\lambda_{\hat{\rho}_{\rm pure}^{(N,N_0)}} (Y) } \, ,
\end{eqnarray}
 given in terms of the single non--zero eigenvalue of the density matrix, is zero. This is natural since it is a measure of the amount of uncertainty or lack of information associated to a quantum state. 

In order to have a {\sl mixed state} multiple non--zero eigenvalues must be present in the spectrum of the density matrix. The effective dimension of a mixed state is defined as the inverse of its purity,
$d^{\rm eff} (\hat{\rho}^{(N)} ) = ({\rm Tr} (\hat{\rho}^{(N)} (\eta))^2 )^{-1}$
and provides a measure of how many pure states contribute significantly to the mixture. 

In the next section we focus on how to modify this picture when non--linear higher--order corrections are introduced in the formalism. This is a very complicated problem if treated in full generality. It is nevertheless possible 
to study how one of its main effects, the suppression of infrared components, affects the hermeticity and spectrum of eigenvalues of the BFKL Hamiltonian in the theoretical framework under discussion in this work.

\section{Screening of infrared diffusion}
\label{sec:scre-infr-diff}

An important consequence of introducing the interaction of the BFKL Pomeron with multiple reggeized gluon states,  is 
the suppression of diffusion into low virtualities (see, {\it e.g}.\cite{Mueller:2002zm, Bartels:2007dm}). In order to investigate the implication of this effect in 
the quantum properties of the BFKL states we will study a modification of the original Hamiltonian 
in Eq.~\ref{BFKLeqn1} which has been already investigated in~\cite{BethencourtdeLeon:2011xks}. This amounts to introducing an asymmetry between infrared and ultraviolet diffusion by suppressing the former in the form 
\begin{eqnarray}
(\hat{\mathcal{H}}^{{\rm dressed}}_N)_{i, j} &=& \sum_{n=1}^{N-1} 
\left(\frac{j}{i}\right)^\kappa
\frac{\delta_i^{j+n}}{n}+\sum_{n=1}^{N-1} \frac{\delta_{i+n}^j}{n}-2 h(i-1) \delta_i^j \nonumber\\
&=& \left(\frac{j}{i}\right)^\kappa \frac{\theta(i-j)}{i -j} + \frac{\theta(j-i)}{j -i} - 2 h(i-1) \delta_i^j\, .
\label{IRDRHam}
\end{eqnarray}
We will study in which range of the real--valued positive parameter $\kappa$ it is possible to operate with a proper quantum state in any region of the coupling constant complex plane.

\begin{figure}
\center
\vspace{-6cm}
  \includegraphics[width=12.cm]{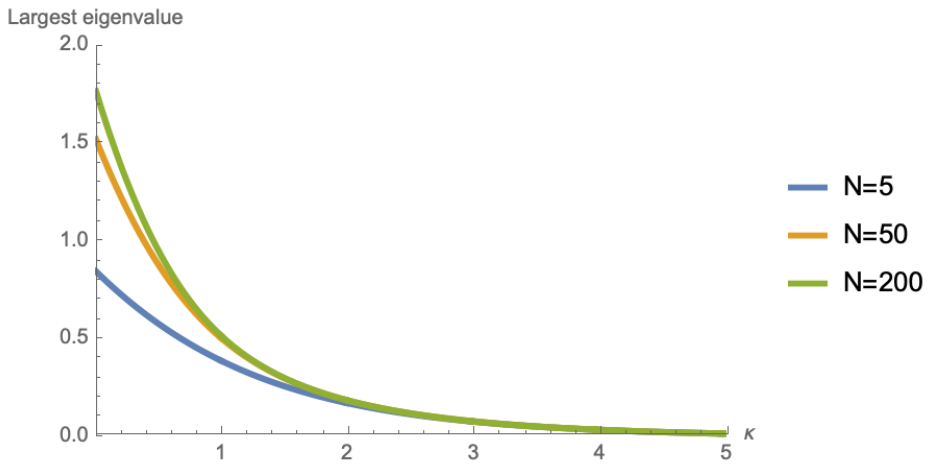}
  \vspace{-6cm}
  \caption{Largest eigenvalue for Hamiltonians with different $N$ as a function of $\kappa$. }
  \label{fig:LargeEigenvalueN5}
\end{figure}

The infrared--dressed Hamiltonian~\ref{IRDRHam} is no longer Hermitian. This has an important effect on the spectrum of the theory. It still consists of real eigenvalues where the largest, positive, one gets rapidly reduced as $\kappa$ increases~\cite{BethencourtdeLeon:2011xks}. This is easy to understand since as $\kappa \to \infty$ the Hamiltonian  becomes a triangular matrix whose eigenvalues correspond to the diagonal elements $-2 h (i-1)$, with $i=1,2, \dots$. This is shown for $N=5, 50, 200$ in Fig.~\ref{fig:LargeEigenvalueN5}.
For example, for $\kappa=0.5, N=5$ the original spectrum  is modified to $(-4.92985,-4.00719,-2.94808,-1.52508,0.576859)$. The associated Green's function state for the initial condition of 
Eq.~\ref{InitiCond3} reads
\begin{eqnarray}
\left|{\phi}^{(5)} \right> \simeq 
\left(
\begin{array}{ccccc}
 0.28 &-0.34&0.04&0.01&0.00\\
 0.11&0.30 &-0.25 &-0.16 &0.00 \\
 0.06&0.26&0.10 &0.55 &0.03  \\
0.04& 0.18 &0.15 &-0.26 &-0.10 \\
0.02& 0.11 &0.10 &-0.34 &0.10 \\
\end{array}
\right)
\left(\begin{array}{c}
e^{0.58 \eta} \\
e^{  -1.53\eta}\\
e^{ -2.95\eta } \\
 e^{-4.01 \eta } \\
  e^{-4.93 \eta} \\
\end{array}\right).
 \end{eqnarray} 
 
\begin{figure}
\vspace{-5cm}
\center
  \includegraphics[width=12.cm]{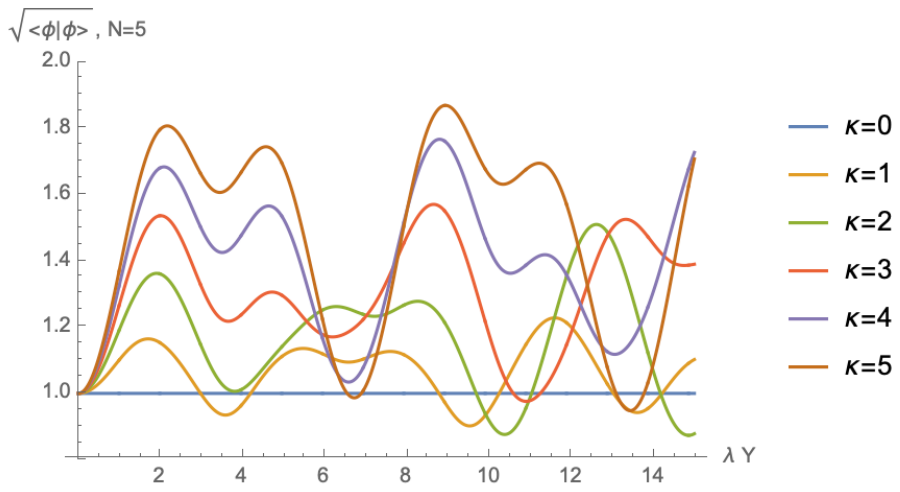}
  \vspace{-6cm}
  \caption{Normalization of the BFKL quantum state along the line $\alpha = i \lambda$ for the coupling. It scales with the product $\lambda Y$. The flat line corresponds to ${\cal H}^{\Box }_5$ while the oscillatory ones corresponds to  $\hat{\cal H}^{\rm dressed}_5$ with different values of the screening parameter $\kappa$.}
  \label{fig:NormalizationPhi}
\end{figure}

 As we have already discussed, when $\kappa=0$ the norm of the BFKL state along the line 
 $\alpha = i \lambda$ is one for any $Y$. Even if it is non--Hermitian, the dressed Hamiltonian does not generate complex eigenvalues  although its eigenvectors do not form an orthogonal set. This implies that the quantum state is no longer normalized to unity. In Fig.~\ref{fig:NormalizationPhi} this normalization is shown for $N=5$ and different values of the screening parameter $\kappa$. With oscillations in $\lambda Y$, the scaling variable, the norm is larger than one for non--zero values of 
 $\kappa$. 
It is possible to interpret this change in the normalization of the state as a consequence of the interaction with  diagrams containing multiple reggeized gluon states.  Their influence in the system is parametrized  by $\kappa$. A study where this idea is put forward can be found in Ref.~\cite{Bartels:2007dm} where it is shown that in non-linear evolution equations in the zero conformal spin sector (as the one considered here) the Green's functions receives the bulk of the contributions from anti--collinear configurations where the infrared/ultraviolet symmetry is manifestly broken. 

\begin{figure}
\center
  \includegraphics[width=6.cm]{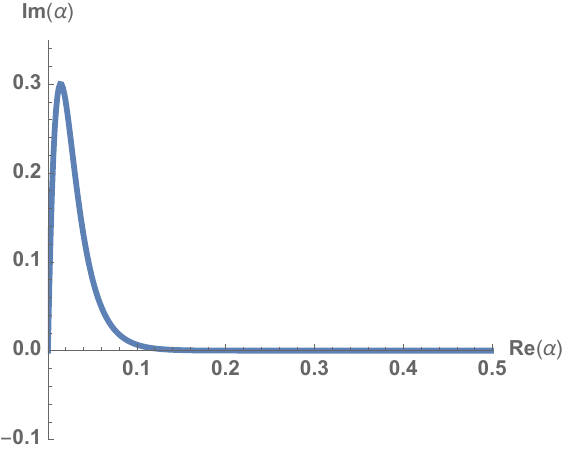}
  \caption{Analytic continuation path for the coupling  
  $\alpha = \lambda + i \, 
e^{-\sigma \lambda} \tanh(\sigma \lambda) $ with $\sigma=50$.}
  \label{fig:CouplingComplex}
\end{figure}

To study  the quantum state in the physical region it is needed to analytically continue  to the  real line for $\alpha$. For this the path $\alpha = \lambda + i \, e^{-\sigma \lambda} \tanh(\sigma \lambda) $ with $\sigma=50$ is chosen (see Fig.~\ref{fig:CouplingComplex}). This particular choice is arbitrary but it allows us to transit from a region of bounded normalization of the quantum state, very close to the imaginary axis, towards the physical region of real coupling smoothly,  while keeping the modulus of the coupling small. Our conclusions are independent of the choice of path (as they also are of the size of $N$). 

\begin{figure}
\vspace{-8cm}
\center
  \includegraphics[width=12.cm]{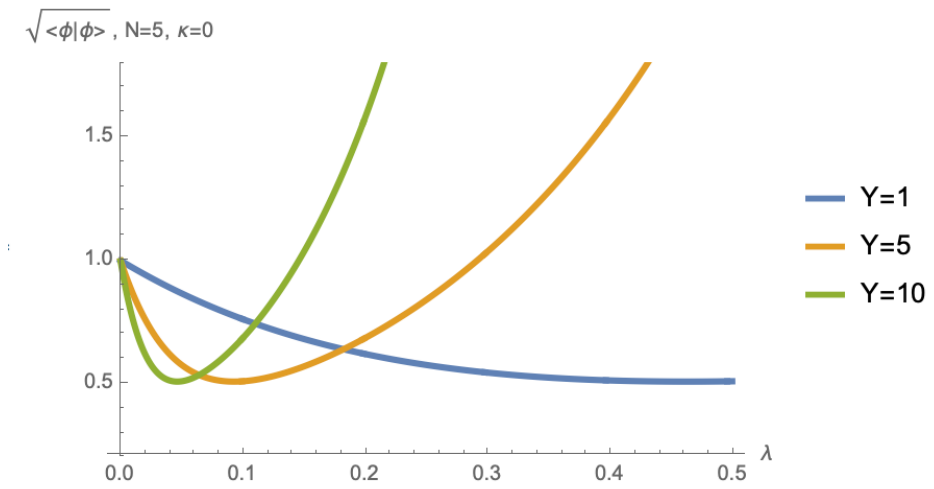}\\
  \vspace{-11.4cm} 
  \includegraphics[width=12.cm]{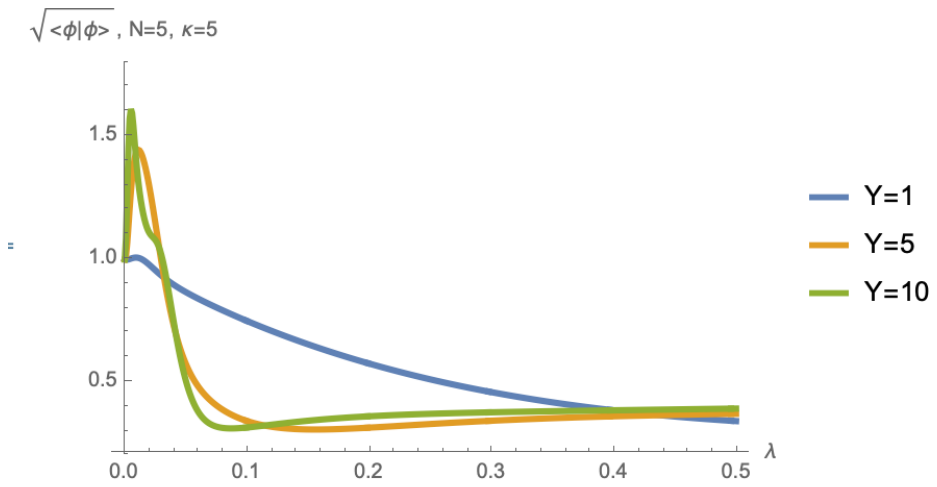}\\
  \vspace{-6cm}
  \caption{Normalization of the BFKL quantum state as a function of $\lambda$ with coupling $\alpha = \lambda + i \, 
e^{-\sigma \lambda} \tanh(\sigma \lambda) $, $\sigma=50$, for different values of $Y$ and $N=5$, in the original formulation (top) and with infrared suppression (down).}
  \label{fig:NormalizationComplexPath}
\end{figure}

\begin{figure}
\vspace{-7cm}
\center
  \includegraphics[width=12.cm]{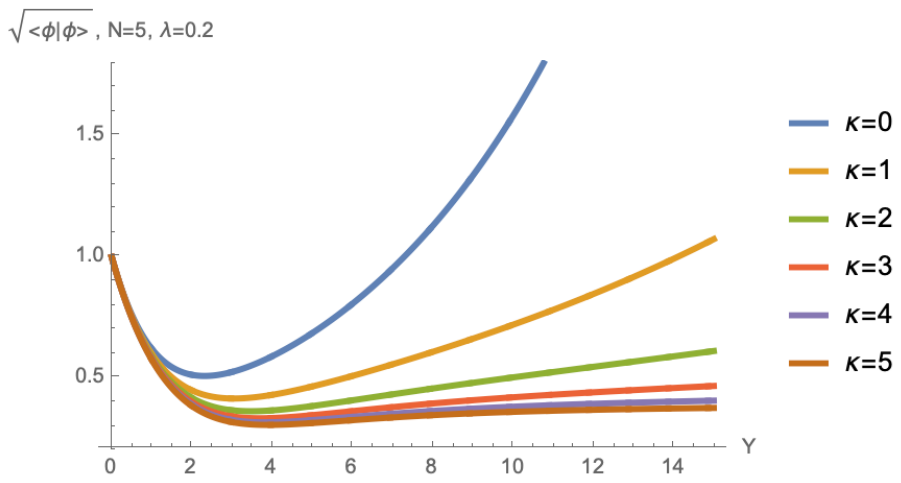} \\
  \vspace{-12cm}
  \includegraphics[width=12.cm]{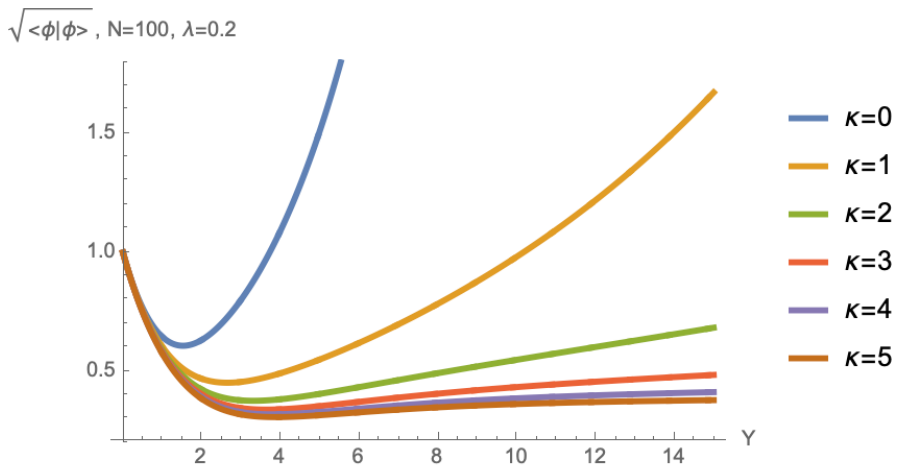}
  \vspace{-6cm}
  \caption{Normalization of the quantum state with infrared dressing $\kappa$ 
  as a function of $Y$ with coupling $\alpha \simeq \lambda = 0.2$. $N=5$ (top) and $N=100$ (down).}
  \label{fig:NormalizationvsYlambda02}
\end{figure}

 As expected from the generic properties of the BFKL Pomeron,  a fast 
rise of the norm of the BFKL quantum state appears when the system approaches the region of physical coupling. This effect is larger for larger values of $Y$ and is plotted in Fig.~\ref{fig:NormalizationComplexPath} (top).
This drastically changes when introducing the infrared screening as can be seen in Fig.~\ref{fig:NormalizationComplexPath} (down). In this case the normalization of the state is smaller than one for any value of the coupling near the real line and decreases as the energy increases. In Fig.~\ref{fig:NormalizationvsYlambda02} (top) it is shown, for $\alpha \simeq 0.2, N=5$ and increasing values of $\kappa$, how the infrared dressing removes a large fraction of the probability associated to the state as $Y$ increases. The infrared suppression saturates its effect at $\kappa \sim {\cal O}(5)$. This final state configuration at larger $\kappa$ is invariant with $N$, see Fig.~\ref{fig:NormalizationvsYlambda02} (down) for $N=100$.  

\begin{figure}

\vspace{-6.5cm}
\hspace{-3.cm}  \includegraphics[width=11.cm,angle=0]{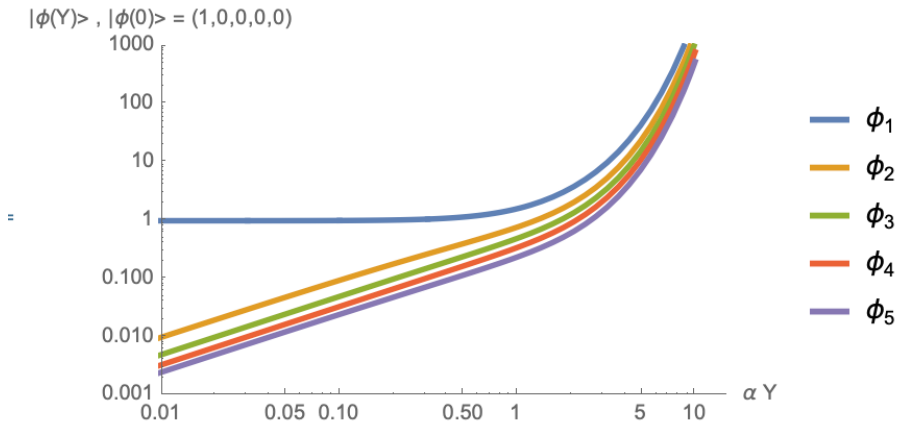} \hspace{-3.cm} \includegraphics[width=11.cm,angle=0]{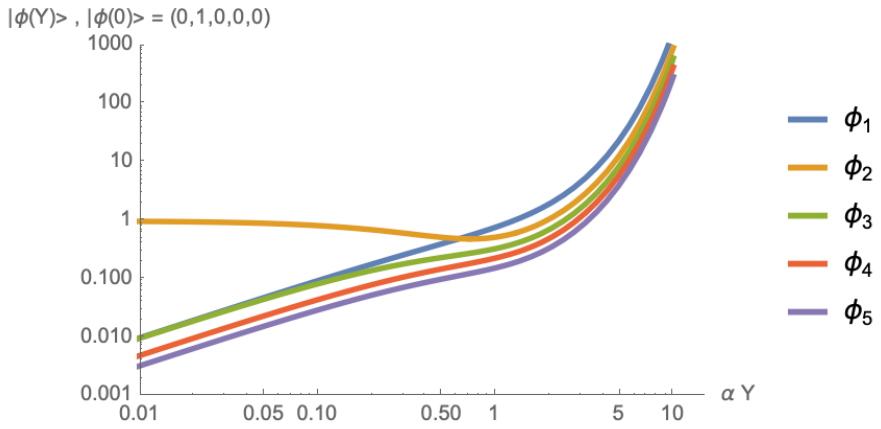} 

\vspace{-11.5cm}

\hspace{-3.cm}  \includegraphics[width=11.cm,angle=0]{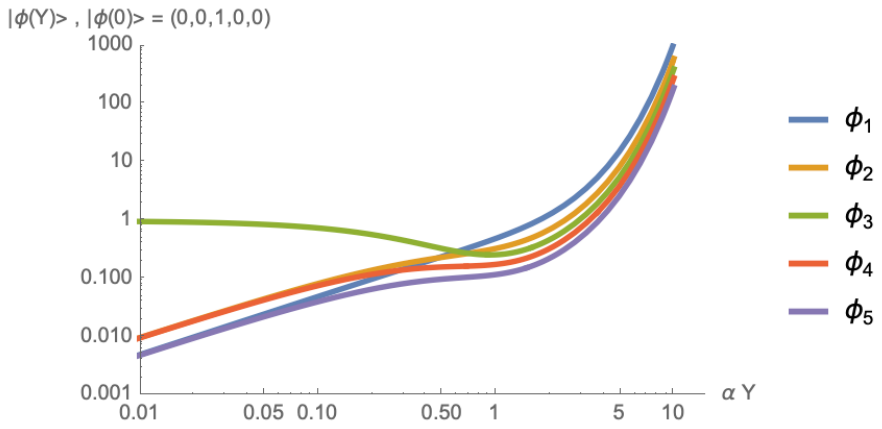} \hspace{-3.cm} \includegraphics[width=11.cm,angle=0]{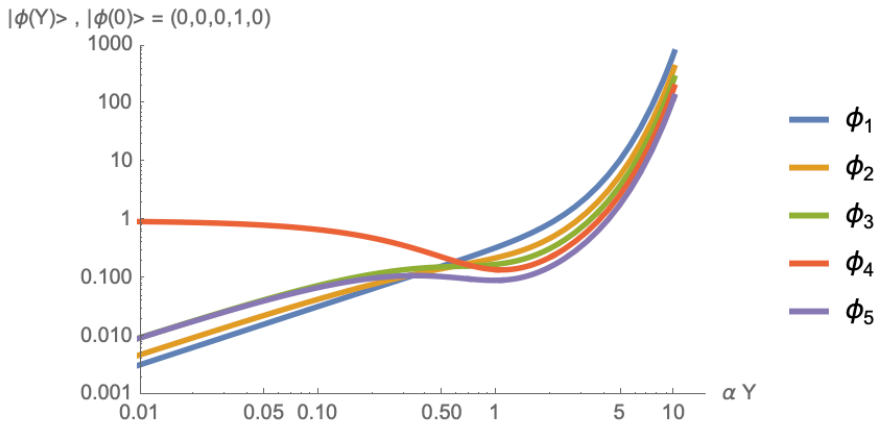}

\vspace{-11.5cm}

\hspace{-3.1cm}    \includegraphics[width=11.cm,angle=0]{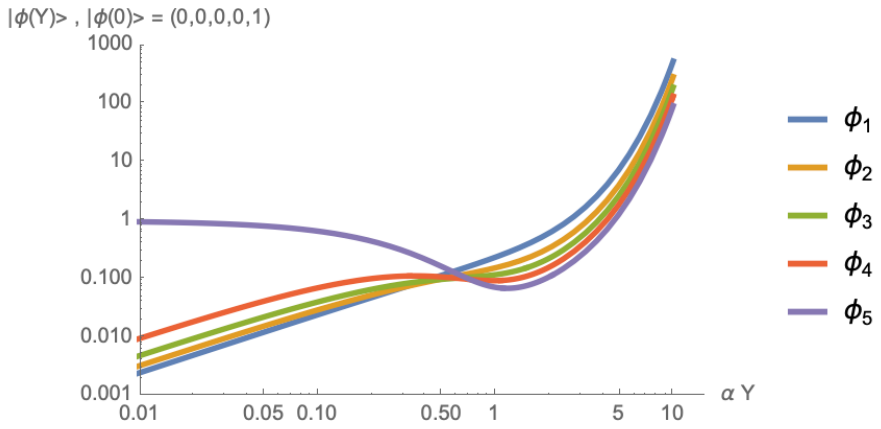} 

\vspace{-5.5cm}

  \caption{Original BFKL evolution for $N=5$ with different initial pure states.  }
  \label{fig:N5PsUno}
\end{figure}

\begin{figure}
\vspace{-9cm}
\hspace{-3.cm}  \includegraphics[width=11.cm,angle=0]{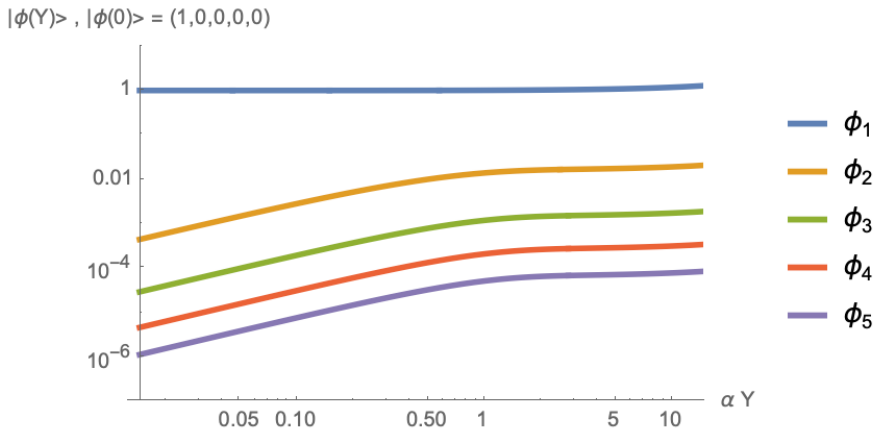} \hspace{-3.cm} \includegraphics[width=11.cm,angle=0]{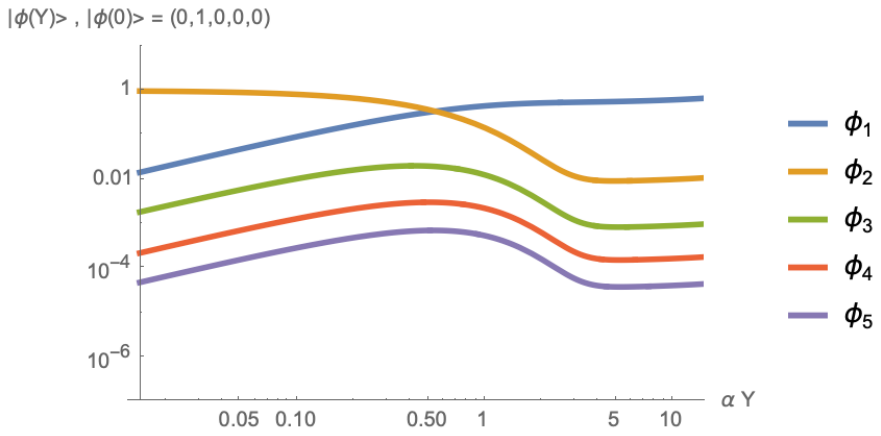} 

\vspace{-11.5cm}

\hspace{-3.cm}  \includegraphics[width=11.cm,angle=0]{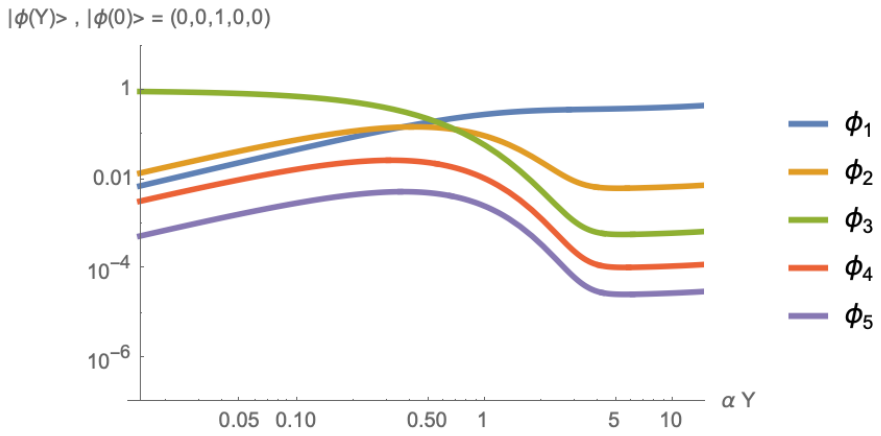} \hspace{-3.cm} \includegraphics[width=11.cm,angle=0]{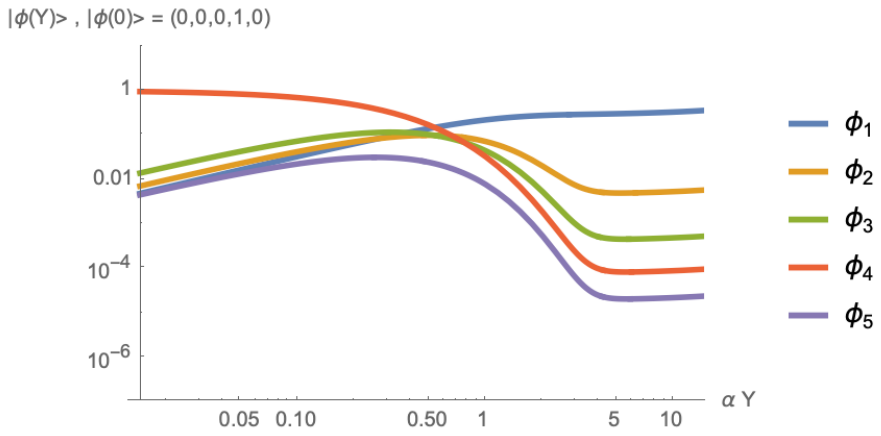}

\vspace{-11.5cm}

\hspace{-3.1cm}   \includegraphics[width=11.cm,angle=0]{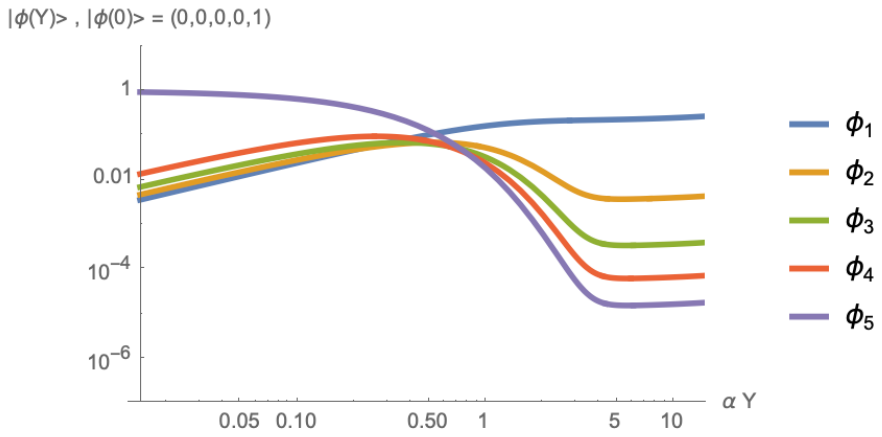} 

\vspace{-5.5cm}
  \caption{Five solutions of the $N=5$ system with different initial conditions as pure states when the largest eigenvalue is suppressed by the dressing operator.  }
  \label{fig:kappasSteady}
\end{figure}

Along the real line for the coupling, and upon evolution, the original system (with $\kappa=0$) rearranges itself from the initial pure state at $Y=0$ into the asymptotic configuration. After a finite amount of evolution in energy the different virtuality components of the quantum state converge to the same stable configuration. This is seen in Fig.~\ref{fig:N5PsUno} where five distinct pure state initial conditions, exciting different virtuality modes, are plotted. It can be seen how the five (for $N=5$) component state loses ``memory" of the initial condition very rapidly. 
For real coupling and sufficiently large $\kappa$, the rapidity evolution of the quantum state can become stable in a given range of $Y$. An example is Fig.~\ref{fig:kappasSteady} where, for $N=5$ and $\kappa=5$, we see how  all virtuality modes reach a flat behaviour at large values of $Y$. 

Non--Hermitian Hamiltonians appear in 
open quantum systems where some sort of interaction with an external environment is present. In the BFKL system this external actor would be the higher--order non--linear quantum corrections. In the next section we investigate the possibility of generating decoherence effects due to the suppression of infrared modes and how this is related to the normalization of the quantum state. We will also show how to regain a probabilistic picture within this setup.

\section{Non--hermitian Hamiltonian}
\label{sec:non-herm-hamilt}

\begin{figure}
\vspace{-7cm}
  \includegraphics[width=12.cm]{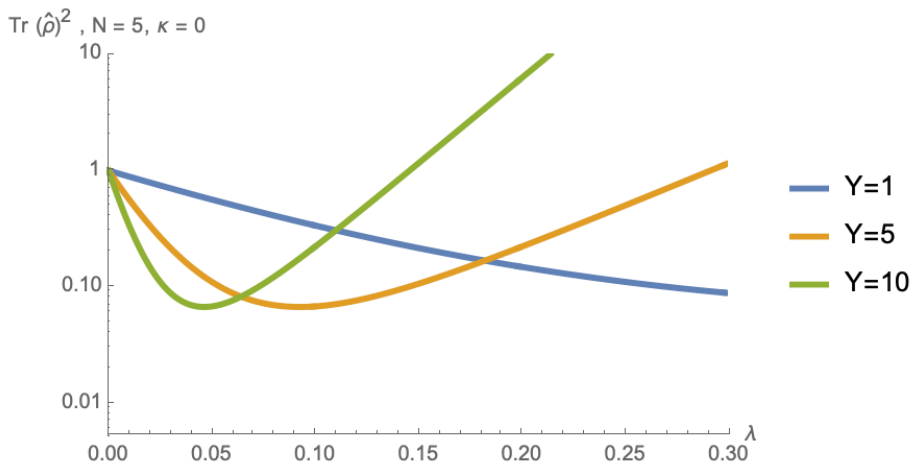}\\
  \vspace{-12.4cm}\\
  \includegraphics[width=12.cm]{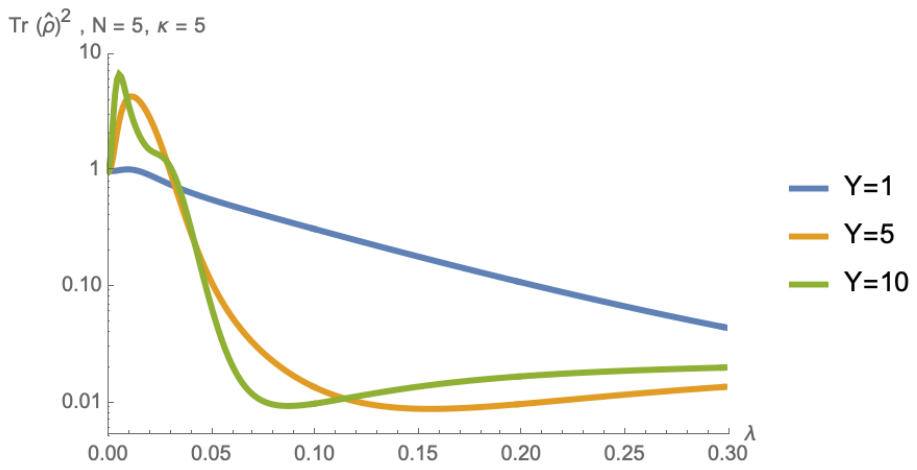}
  \vspace{-6cm}
  \caption{Purity of the BFKL quantum state as a function of $\lambda$ with coupling $\alpha = \lambda + i \, 
e^{-\kappa \lambda} \tanh(\kappa \lambda) $, $\kappa=50$, for different values of $Y$, in the original formulation (top) and with infrared suppression (down).}
  \label{fig:PurityBFKLToRealLineY}
\end{figure}

As we have seen, the original formulation of the BFKL Hamiltonian along the line $\alpha = i \lambda$ transforms pure states into pure states since the trace of the associated density matrix, which is a projector, is one. This cannot be the case for the infrared dressed Hamiltonian since the normalization of the state changes with the variable $\lambda Y$, Fig.~\ref{fig:NormalizationPhi}. This implies that the trace and purity of the density matrix grow to values bigger than one and hence spoil the quantum properties of the system. It is therefore mandatory to return to the physical region for the coupling to investigate how pure states evolve into mixed ones generated by the decoherence process introduced through the infrared dressing, considered as an effective description of higher--order quantum non--linearities. If this is done for 
the original BFKL formalism, Fig.~\ref{fig:PurityBFKLToRealLineY} (top), instabilities in the purity of the state soon appear when $Y$ and the value of the coupling are large enough. The structure of the quantum state is not 
properly defined especially when being very close to physical values of the 
coupling. 

\begin{figure}
\vspace{-7.9cm}
\center
  \includegraphics[width=12.cm]{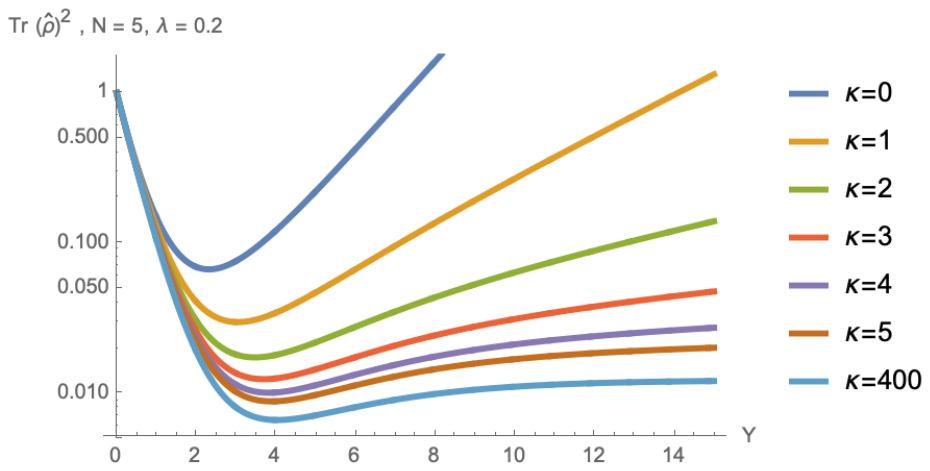}\\
  \vspace{-12cm}
  \includegraphics[width=12.cm]{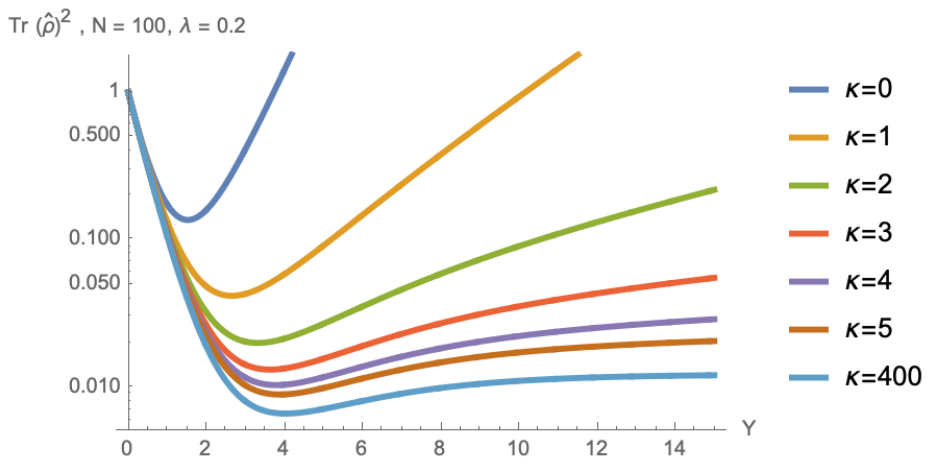}
  \vspace{-6cm}
    \caption{Purity of the BFKL quantum state as a function of $Y$ with coupling $\alpha \simeq \lambda = 0.2$ in the original formulation and with infrared suppression ($\kappa \neq 0$). $N=5$ (top), $N=100$ (down)}
  \label{fig:PurityBFKLToRealLineYIRalpha02}
\end{figure}

This situation largely improves if the infrared dressing is implemented. In Fig.~\ref{fig:PurityBFKLToRealLineY} (down)  it can be seen that for $\hat{\cal H}_N^{\rm dressed}$ a smooth transition emerges for  any non--zero value of $Y$ from a pure state when the coupling tends to zero towards a highly mixed state in its physical region. This process of decoherence takes place at a faster pace as $Y$ rises. If the coupling is fixed to $\alpha \simeq 0.2$ and the $Y$ dependence of the purity is studied, Fig.~\ref{fig:PurityBFKLToRealLineYIRalpha02} (top), a rapid rise for the original 
$\hat{\cal H}^\square_N$ is found. The dressed Hamiltonina, $\hat{\cal H}_N^{\rm dressed}$, leads on the other hand to a rapid transition, 
faster as $\kappa \geq 4$ grows, from the initial pure state  towards a highly--mixed state at large values of $Y$. This picture is not modified as $N$ grows, see Fig.~\ref{fig:PurityBFKLToRealLineYIRalpha02} (down) for $N=100$.  

\begin{figure}
  \vspace{-6cm}
\center
  \includegraphics[width=12.cm]{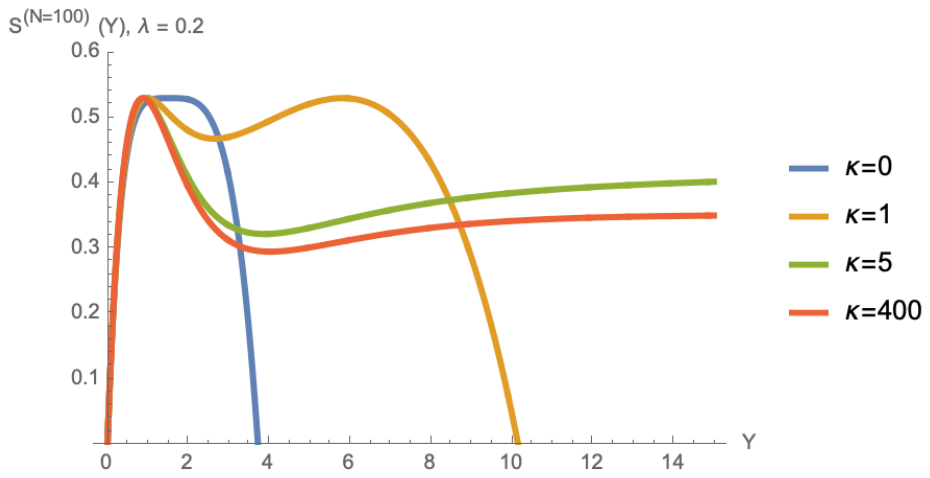}  
  \vspace{-6cm}
  \caption{Von Neumann entropy as a function of $Y$ for coupling $\alpha \simeq \lambda = 0.2$ with infrared suppression.}
  \label{fig:vNEntropyDrop}
\end{figure}

In order to deal with an open system driven by a non--hermitian Hamiltonian such as the dressed hamiltonian in Eq.~\ref{IRDRHam}, it is useful to express it as a sum of symmetric and antisymmetric parts, {\it i.e.}
\begin{eqnarray}
\hat{\cal H}^{\rm dressed}_N &=& \hat{H}^+_N + \hat{H}^-_N\, , \\
\hat{H}^+_N &=& (\hat{H}_N^+)^T ~=~ \frac{1}{2} \left(\hat{\cal H}^{\rm dressed}_N + (\hat{\cal H}^{\rm dressed}_N)^T\right) \, ,\\
\hat{H}^-_N &=& - (\hat{H}^-_N)^T ~=~ \frac{1}{2} \left(\hat{\cal H}^{\rm dressed}_N - (\hat{\cal H}^{\rm dressed}_N)^T\right) \, .
\end{eqnarray}
The evolution with energy of the quantum state is now expressed in terms of these two Hamiltonians. For a coupling of the form $\alpha = \lambda + i f(\lambda)$, with $\lambda, f \in \Re$, it reads
\begin{eqnarray}
 \frac{\partial}{\partial Y} \left|{\phi}^{(N)}\right>
&=&  (\lambda + i f(\lambda))\hat{H}^+_N  \, \left|{\phi}^{(N)} \right>
+ (\lambda + i f(\lambda))\hat{H}^-_N  \, \left|{\phi}^{(N)} \right> \, , \\
\frac{\partial}{ \partial Y}  \left<{\phi}^{(N)}\right|
&=&  (\lambda - i f(\lambda)) \left<{\phi}^{(N)} \right| \,  \hat{H}^+_N   
-(\lambda - i f(\lambda))  \left<{\phi}^{(N)} \right| \, \hat{H}^-_N   \, .
\end{eqnarray}
For the density matrix $\hat{\rho}^{(N)} (Y) = \left|{\phi}^{(N)}  \right>\left<{\phi}^{(N)}\right|$ this implies an evolution driven by commutators and anticommutators,
\begin{eqnarray}
\frac{\partial}{\partial Y}   \hat{\rho}^{(N)}  &=&  
\lambda \left\{\hat{H}^+_N  ,   \hat{\rho}^{(N)} \right\}
+ i f(\lambda) \left[\hat{H}^+_N   ,  \hat{\rho}^{(N)}  \right]    
 \nonumber\\ 
 &+&  
 \lambda \left[\hat{H}^-_N ,   \hat{\rho}^{(N)} \right] 
   + i f(\lambda) \left\{\hat{H}^-_N ,   \hat{\rho}^{(N)}\right\} \, .
\end{eqnarray}
The derivative of its trace is
\begin{eqnarray}
\frac{\partial}{\partial Y}  \,  {\rm Tr} (\hat{\rho}^{(N)} ) 
&=&2 \lambda \, {\rm Tr} (\hat{H}^+_N   \hat{\rho}^{(N)} )   
 \, .
\end{eqnarray}
This translates into the purity ${\rm Tr} ((\hat{\rho}^{(N)} )^2) =  (\sum_{L=1}^N |\phi_L|^2 )^2$  of the quantum state: 
\begin{eqnarray}
\frac{\partial}{\partial Y} {\rm Tr} \, ((\hat{\rho}^{(N)})^2) 
&=&  4 \lambda\, {\rm Tr} (\hat{\rho}^{(N)} )  {\rm Tr} (\hat{H}^+_N   \hat{\rho}^{(N)} )   \, .
\end{eqnarray}
The corresponding von Neumann entropy of the quantum system
 \begin{eqnarray}
 S_{\rm vN}^{(N)} &=&- {\rm Tr} \left(\hat{\rho}^{(N)} 
 \log_2 \hat{\rho}^{(N)} \right) 
 \end{eqnarray}
 is studied in Fig.~\ref{fig:vNEntropyDrop}. It can be seen how a very fast decoherence process takes place at small energy. After reaching a maximum at a small value of $Y$, with an entropy, $S_{\rm vN}^{(N)}  \simeq 0.5$, corresponding to a highly but not completely mixed state, there is a phase of monotonic decrease to  a finite constant value of entropy when $N, Y\to \infty$. The physical origin of this asymptotic finite von Neumann entropy is an interesting source of study for future works. 

To evaluate quantum averages of operators, $\left< {\cal O}\right>_{Y} = {\rm Tr} \left(  {\hat {\cal O}} \, \hat{\Omega}^{(N)}   \right)$, it is needed to use a normalized density matrix with trace one see, {\it e.g.} \cite{Sergi:2014eja}), 
\begin{eqnarray}
\hat{\Omega}^{(N)}  &\equiv& \frac{\hat{\rho}^{(N)} }{{\rm Tr} \left(\hat{\rho}^{(N)}  \right)} \, .
\end{eqnarray}
This is idempotent, since 
${\rm Tr} ((\hat{\rho}^{(N)} )^2) =  
({\rm Tr} (\hat{\rho}^{(N)} ) )^2$. Its energy derivative  follows 
\begin{figure}
\vspace{-6cm}
\center
  \includegraphics[width=12.cm]{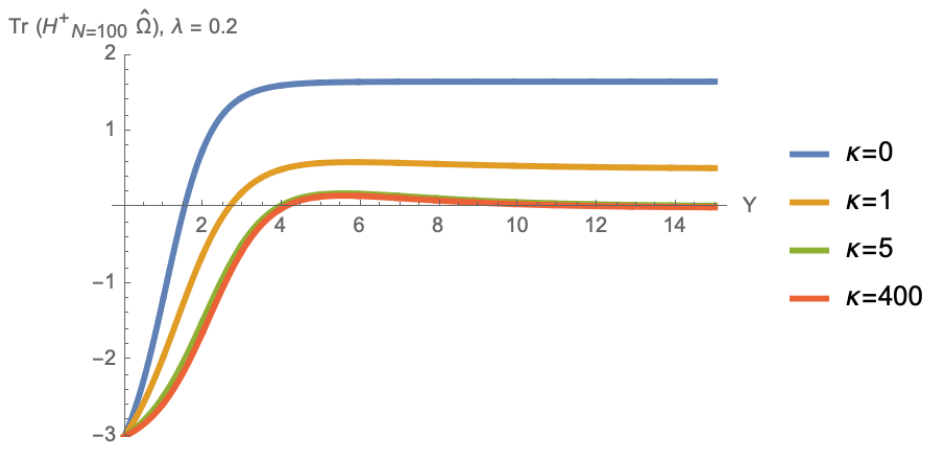}  
   \vspace{-6cm}
  \caption{$ {\rm Tr} \left(\hat{\cal H}^+_N  \,  \hat{\Omega}^{(N)}  \right)$ as a function of $Y$ for coupling $\alpha \simeq \lambda = 0.2$ with infrared suppression.}
  \label{fig:TrHpOmega}
\end{figure}
\begin{eqnarray}
\frac{\partial}{ \partial Y} \hat{\Omega}^{(N)} &=&
 i f(\lambda) \left[\hat{H}^+_N   ,  \hat{\Omega}^{(N)}  \right] 
+ i f(\lambda) \left\{\hat{H}^-_N ,   \hat{\Omega}^{(N)}\right\}
 \nonumber\\
&+& \,\lambda \left\{\hat{H}^+_N  ,   \hat{\Omega}^{(N)} \right\}
+ \lambda \left[\hat{H}^-_N ,   \hat{\Omega}^{(N)} \right] 
   - \, 2 \lambda \, 
 \hat{\Omega}^{(N)}   \left< \hat{H}^+_N\right>_{Y}  \, .
 \label{EvoEqOmega}
\end{eqnarray}
The non--linear last term ensures probability conservation, $\frac{\partial}{\partial Y}
{\rm Tr} (\hat{\Omega}^{(N)}) = 0$. It is interesting to observe that 
$ \left< \hat{H}^+_N\right>_{Y}$ tends to zero in the large $Y$ limit (Fig.~\ref{fig:TrHpOmega}) if the  values of $\kappa$ are sufficiently large  to stabilize the entropy. 

There are very interesting approaches in the literature which have investigated the concept of a density matrix in non--linear evolution equations. A particular one, in the context of the Color Glass Condensate 
is that of Ref.~\cite{Armesto:2019mna}. Although it is a much more sophisticated approach than the one  presented here, they also found a decrease of purity with energy and were able to describe the system with a Linbland equation associated to an open system (it is also known as Franke--Gorini--Kossakowski--Lindblad--Sudarshan equation~\cite{Franke:1976tx,Gorini:1975nb,Lindblad:1975ef}). 

The evolution equation in~\ref{EvoEqOmega} can be written in Lindblad form once $\hat{H}^+_N
= \hat{L}_N^T \hat{L}_N$, with $\hat{L}_N = \hat{{D}}_N \hat{Q}_N $, $\hat{{D}}_N={\rm diag}(\sqrt{\mu_1}, \dots, \sqrt{\mu_N})$,  where $\mu_i$ are the eigenvalues of $\hat{H}^+_N$,  {\it i.e.}
\begin{eqnarray}
{\rm Tr} \left(\hat{H}^+_N  \,  \hat{\Omega}^{(N)}  \right) 
&=& {\rm Tr} \left(  \hat{L}_N  \,  \hat{\Omega}^{(N)}  \hat{L}_N^T\right) \, .
\end{eqnarray}
Finally, the evolution equation reads
\begin{eqnarray}
\frac{\partial}{ \partial Y} \hat{\Omega}^{(N)} &=& 
\left(\lambda + i f(\lambda) \right) \hat{\cal H}^{\rm dressed}_N \, 
  \hat{\Omega}^{(N)} + {\rm h.c.} 
   - \, 2 \lambda \, 
 \hat{\Omega}^{(N)} {\rm Tr} \left(  \hat{L}_N  \,  \hat{\Omega}^{(N)}  \hat{L}_N^T\right)  \, .
\end{eqnarray}
This corresponds to a driven open quantum system with no dissipation but some external fluctuations acting to preserve probability. On the real line, $f=0$, it can also be written as 
\begin{eqnarray}
\frac{\partial}{ \partial Y} \hat{\Omega}^{(N)} &=& 
\lambda  \, \left(\hat{H}^{-}_N + \hat{L}_N^T \hat{L}_N\right) 
  \hat{\Omega}^{(N)} + {\rm h.c.} 
   - \, 2 \lambda \, 
 \hat{\Omega}^{(N)} {\rm Tr} \left(  \hat{L}_N  \,  \hat{\Omega}^{(N)}  \hat{L}_N^T\right)  \, .
 \label{LiOm}
\end{eqnarray}
This equation has quasi--Lindbladian structure where $\hat{H}^{-}_N$ corresponds to coherent evolution of a system inside a quantum environment which generates  dissipation represented by $\hat{L}_N^T \hat{L}_N$, minus quantum fluctuations responsible for probability conservation.  Instead of having $N^2-1$ Lindblad operators as in the standard Lindblad equation there is only one, $\hat{L}_N$. 

The associated  entropy, Fig.~\ref{fig:vNEntropyDrop}, reaches a plateau at a non--zero value for large $Y$. This follows a fast period of Lindblad decoherence. The roles of system and environment are somehow reversed from what one would naively argue. It would be more natural to obtain a picture where the hermitian hamiltonian $\hat{H}^{+}_N$ drives the evolution, receiving corrections encoded in $\hat{H}^{-}_N$, but just the opposite 
appears. It is also noteworthy that the corrections due to quantum fluctuations, after the period of decoherence, generate a small and negative contribution to the derivative of the density matrix (Fig.~\ref{fig:TrHpOmega} times $-2 \lambda \, \hat{\Omega}^{(N)}$). Interpreted as a gain and loss system we find that the environment provides the latter while the non--hermitian contribution provides a source of the former where now the trace of the density matrix is conserved.

\section{Conclusions}
\label{sec:conclusions}

 The resummation of high energy logarithms present in  scattering amplitudes in QCD and supersymmetric theories leads to the BFKL equation. By discretizing the space of virtualities in loop corrections and regularizing it inside a box, it is possible to explore quantum properties of the state associated to the gluon Green's function. For imaginary values of the coupling its normalization is bounded and preserved under rapidity evolution. This implies that pure states of well defined virtuality are evolved into pure states characterized by a density matrix with purity one. 
 
 As a model of non--linear corrections, which introduce contributions to unitarity via suppression of the diffusion intro infrared modes, we study the spectrum of a modified Hamiltonian whose largest positive eigenvalue can become arbitrarily small. This introduces quantum decoherence which affects the normalization of the state forcing to analytically continue the coupling to the real line. This shows how the rapidity evolution of the state generates unbounded normalization for usual BFKL while a bounded one for enough saturation of infrared modes. This implies that to obtain correct quantum properties at high energies such as a purity smaller than one or a positive von--Neumann entropy it is needed to break the infrared/ultraviolet original symmetry of the BFKL equation. 
 
Stemming from a non--hermitian Hamiltonian, the density matrix describing the open system, when normalized, fulfils an evolution equation of Lindblad type with dissipation and quantum fluctuations, Eq.~\ref{LiOm}.  Much work remains to be done for the future. This includes the introduction of higher order corrections both in QCD and supersymmetric theories and the implementation of a more precise model of unitarization corrections to allow for a more complete comparison to previous works where the Lindblad equation of an open system appears~\cite{Armesto:2019mna,Li:2020bys}. Although the results here presented are robust, it is desirable to connect with other approaches in forthcoming works.

\section*{Acknowledgements}

The work of G.C. was supported by the Funda\c{c}{\~ a}o para a Ci{\^ e}ncia e a Tecnologia (Portugal) under project CERN/FIS-PAR/0032/2021 and contract ‘Investigador FCT - Individual Call/03216/2017’ and by project EXPL/FIS-PAR/1195/2021. M.H. would like to thank the IFT UAM/CSIC for hospitality.  The work of A.S.V.  is partially supported by the Spanish Research Agency (Agencia Estatal de Investigaci{\'o}n) through the Grant IFT Centro de Excelencia Severo Ochoa No CEX2020-001007-S, funded by MCIN/AEI/10.13039/501100011033 and the Spanish Ministry of Science and Innovation grant PID2019-110058GB-C21/ C22. It has also received funding from the European Union’s Horizon 2020 research and innovation programme under grant agreement No. 824093.

\end{document}